%% file: cdc2020.tex
\documentclass[letterpaper, 10 pt, conference]{ieeeconf}
\IEEEoverridecommandlockouts

\usepackage{cite}
\usepackage{amsmath,amssymb,gensymb, bm, amsfonts, mathtools}
\usepackage{algorithmic}
\usepackage{graphicx}
\usepackage{textcomp}
\usepackage{xcolor}
\usepackage{xtab}
\usepackage{balance}
\include{macros}

\include{notation_defs}

\newenvironment{audit}{\color{orange}}{\color{black}}
\newcommand{\todo}[1]{{\color{red} #1}}

\renewcommand{\todo}[1]{}

\newtheorem{theorem}{Theorem}[section]
\newtheorem{lemma}[theorem]{Lemma}

\title{A Minimum Energy Filter for Localisation of an Unmanned Aerial Vehicle
}

\author{Jack Henderson,  Mohammad Zamani, Robert Mahony, and Jochen Trumpf%
\thanks{
		Jack Henderson, Robert Mahony, and Jochen Trumpf are with Systems Theory and Robotics at the Australian National University, \texttt{firstname.lastname@anu.edu.au}.	Mohammad Zamani is with the Land Division, Defence Science and Technology Group, Australia, \texttt{mohammad.zamani@dst.defence.gov.au}}
\thanks{This research is supported by the Commonwealth of Australia as represented by the Defence Science and Technology Group of the Department of Defence and by the Australian Research Council Discovery Project DP190103615: ``Control of Network Systems with Signed Dynamical Interconnections''}
	}
\begin{document}
\maketitle

\begin{abstract}
Accurate localisation of unmanned aerial vehicles is vital for the next generation of automation tasks.
This paper proposes a minimum energy filter for velocity-aided pose estimation on the extended special Euclidean group.
The approach taken exploits the Lie-group symmetry of the problem to combine Inertial Measurement Unit (IMU) sensor output with landmark measurements into a robust and high performance state estimate.
We propose an asynchronous discrete-time implementation to fuse high bandwidth IMU with low bandwidth discrete-time landmark measurements typical of real-world scenarios.
The filter's performance is demonstrated by simulation.
\end{abstract}


\section{Introduction}
Unmanned aerial vehicles (UAV) are an increasingly important technology in modern society with applications in photography, logistical deliveries, mapping, inspection tasks, etc.
Accurate estimation of a vehicle's position and orientation (pose) is critical for the new generation of applications of such vehicles.
Most pose estimation algorithms for aerial vehicles depend heavily on Global Navigation Satellite Systems (GNSS) to provide ground truth on position \cite{grovesPrinciplesGNSSInertial2008}.
However, GNSS signals become unreliable and fail entirely in urban canyons, forest environments, or any situation where direct reception of the satellite signal is compromised \cite{grovesPrinciplesGNSSInertial2008}.
This has focused attention on the development of algorithms that use exteroceptive sensors such as vision and lidar to provide position information \cite{scaramuzzaVisualOdometryTutorial2011,brossardInvariantKalmanFiltering2018,mahonyGeometricNonlinearObserver2017,vangoorEquivariantObserverDesign2019}.

Historically, the most successful state estimation algorithms for UAV systems have been derived from stochastic principles.
Early work on attitude estimation built upon the Extended Kalman Filter (EKF), typically representing the system state using Euler angles \cite{crassidisSurveyNonlinearAttitude2007}.
Markley's Multiplicative EKF (MEKF) \cite{markleyAttitudeErrorRepresentations2003} introduced a quaternion attitude representation and exploited group multiplication to propagate error estimates.
Recently, Filipe \etal \cite{filipeExtendedKalmanFilter2015} extended this to pose estimation.
Bonnabel \etal \cite{bonnabelLeftinvariantExtendedKalman2007} proposed a general filtering framework for Lie-groups, introducing the Invariant Extended Kalman Filter (IEKF) \cite{barrauInvariantExtendedKalman2017}.
This framework has lead to applications in velocity aided attitude estimation \cite{bonnableInvariantExtendedKalman2009}, visual odometry and Simultaneous Localisation and Mapping (SLAM)
\cite{barrauEKFSLAMAlgorithmConsistency2016,brossardInvariantKalmanFiltering2018}.
An alternative to the stochastic approach is to consider deterministic observer construction or minimum energy filtering.
The complementary filter \cite{mahonyObserversKinematicSystems2013} on the special orthogonal group $\SO(3)$ \cite{mahonyNonlinearComplementaryFilters2008,bonnabelSymmetryPreservingObservers2008}
was fundamental in overcoming the limitations of low quality IMU systems in the early years of aerial robotics.
Pose estimation was studied at the same time
\cite{baldwinNonlinearObserverDOF2009,baldwinComplementaryFilterDesign2007, huaObserverDesignSpecial2011, vasconcelosNonlinearPositionAttitude2010} and there is recent work in this area on SLAM \cite{mahonyGeometricNonlinearObserver2017,vangoorEquivariantObserverDesign2019}.
The deterministic signal perspective also underlies filters based on the principle of minimum energy filtering \cite{mortensenMaximumlikelihoodRecursiveNonlinear1968,hijabMinimumEnergyEstimation1980}.
Minimum energy filters have been developed on the special orthogonal and special Euclidean Lie-groups for attitude and pose estimation problems \cite{zamaniDeterministicAttitudePose2013, zamaniDiscreteUpdatePose2019, bergerSecondOrderMinimum2015}.
Saccon \etal \cite{sacconSecondOrderOptimalMinimumEnergyFilters2016} developed a general minimum energy filtering theory which also applies to second order systems.
Minimum energy filtering combines the advantages of the gain tuning characteristics of stochastic filters with the deterministic geometric insight of the underlying symmetry group, in particular allowing for a non-zero curvature of the associated Lie-group to be chosen.

In this paper, we specialise the general filter described in Saccon \etal \cite{sacconSecondOrderOptimalMinimumEnergyFilters2016} to derive a minimum energy filter to estimate the pose and linear velocity of an unmanned aerial vehicle equipped with IMU and landmark measurements.
The paper extends earlier work that required linear velocity measurements directly \cite{zamaniCollaborativePoseFiltering2019} to a second order model based on the geometry introduced in \cite{bonnableInvariantExtendedKalman2009}.
This Lie-group is known as the extended special Euclidean group $\SE_2(3)$ \cite{barrauEKFSLAMAlgorithmConsistency2016} and provides a model of the pose (attitude and position) along with the linear (but not angular) velocity of the vehicle.
We model the uncertainty in the signals as a deterministic input disturbance and minimize a deterministic least squares cost functional in these disturbance inputs over trajectories that are compatible with the measured outputs.
A key contribution is to incorporate the landmark observations by an  asynchronous update to the filter state that occurs when landmark information is available at discrete times.
The proposed update is based on decomposing the Riccati equation into a part that is dependent on the landmark measurements and a part that models the state evolution.
The landmark information is modelled as Dirac delta signal leading to a discrete update analogous to that obtained in \cite{zamaniDiscreteUpdatePose2019}.
We provide an explicit matrix representation of the filter and a simple simulation to demonstrate the performance of the proposed algorithm, although the principle contribution of the paper is theoretical.

The remainder of the paper is organised as follows.
Section~\ref{sec:preliminaries} introduces the required notation and concepts from differential geometry.
Section~\ref{sec:problem_formulation} formally defines the system model and optimisation problem.
Section~\ref{sec:derivation} introduces the abstract filter derived by Saccon \etal
An explicit matrix representation for the minimum energy filter is presented in Section \ref{sec:explicit_filter}.
Section~\ref{sec:discretisation} provides the proposed discretisation of the
differential equations to deal with asynchronous landmark updates.
A simple simulation is presented in Section~\ref{sec:simulations} before conclusions are given in Section \ref{sec:conclusion}.

\section{Preliminaries and Notation}
\label{sec:preliminaries}
This section introduces the notation and conventions used throughout the paper.

The following symbols related to differential geometry will be used in the same way as in \cite{sacconSecondOrderOptimalMinimumEnergyFilters2016};
\\
\begin{xtabular}{lp{0.55\columnwidth}}
	$G$ & a connected Lie group \\
	$h,g$ & elements of $G$\\
	$\g$ & the Lie algebra associated with $G$\\
	$X,Y$ & elements of $\g$\\
	$[~.~,~.~]$	& the Lie bracket operator \\
	$\g^*$ & the dual of the Lie algebra $\g$\\
	$\mu$ & an element of $\g^*$\\
	$L_g : G \rightarrow G$ & left translation $L_gh = gh$ \\
	$T_h L_g$ & the tangent map of $L_g$ at $h \in G$ \\
	$gX$ & shorthand for $T_e L_g(X) \in T_gG$ \\
	$\langle ~.~ , ~.~ \rangle $ & duality pairing $\langle \mu, X \rangle = \mu(X)$\\
	$.^* $ & adjoint operator, \\
	& $\langle \mu, A X \rangle = \langle A^* \mu, X \rangle$\\
	$V$ & finite dimensional vector space \\
	$f : G \rightarrow V$ & differentiable map\\
	$\der f(g): T_gG \rightarrow V$ & differential of $f$ at $g$\\
	$\der_1$ & differential with respect to the first argument\\
	$\nabla_{\bm{X}}\bm{Y}$ & covariant derivative on vector fields $\bm{X}$ and $\bm{Y}$\\
	$\omega :  \g \times  \g \rightarrow  \g$ & connection function associated with $\nabla$\\
	$\omega_X : \g \rightarrow \g$ & $\omega_X(Y) = \omega(X, Y)$\\
	$T :  \g \times  \g \rightarrow  \g $ &torsion function associated with $\omega$\\
	$T_X :\g \rightarrow \g$ & $T_X(Y) = T(X, Y)$\\
	$\Hess f(g)$ & Hessian operator of a twice-differentiable function $f$
\end{xtabular}

\subsection{Exponential Functor}
Given a linear map $\phi: U \rightarrow V$ and a third vector space $W$, the exponential functor $(.)^W$ lifts the map $\phi$ to the linear map $\phi^W$ : $\mathfrak{L}(W, U) \rightarrow \mathfrak{L}(W, V)$ defined by $\phi^W(\xi) = \phi \circ \xi$

If the reader is unfamiliar with the exponential functor, they should note its appearance in the abstract operator $E(t)$ in \eqref{eq:E(t)_operator}. In Lemma \ref{lem:E(t)} we remove the exponential functor by using the identity
\begin{align}
\left[ \phi^W \circ \Hess f(g) \right] \circ gX = \phi \circ \left[ \Hess f(g) \circ gX \right].
\label{eq:exponential_functor}
\end{align}

\subsection{The Lie Group $\SE_2(3)$}

The filter presented in this paper estimates the position, orientation and linear velocity of a robot, which can conveniently be represented as a matrix Lie group. Here, we define the structure of the extended special Euclidean group, $\SE_2(3)$, and the corresponding Lie algebra, $\se_2(3)$, as in \cite[A.1.2]{barrauEKFSLAMAlgorithmConsistency2016},
\begin{align}
\SE_2(3) &:  \left\{\begin{bmatrix}R & v & x \\ \bm{0} & 1 & 0 \\ \bm{0} & 0 & 1\end{bmatrix} : R \in \SO(3),~ v, x \in \Real{3}\right\},
\\
\se_2(3) &: \left\{ \xi = \begin{bmatrix}\begin{matrix} (\xi_R)_\times & \xi_v & \xi_x \end{matrix} \\ \bm{0}_{2\times 5}\end{bmatrix} : \xi_R, \xi_v, \xi_x \in \Real{3} \right\}.
\label{eq:def_se_2(3)}
\end{align}

The skew operator, $(.)_\times$, transforms a vector to a skew-symmetric matrix;

\begin{align}
(.)_\times &: \Real{3} \rightarrow \so(3),
&
&\omega_\times := \begin{bmatrix} 0 & -\omega_3 & \omega_2 \\ \omega_3 & 0 & -\omega_1 \\ -\omega_2 & \omega_1 & 0 \end{bmatrix},
\\
\vex &: \so(3) \rightarrow \Real{3},
&
&\vex(\Omega) := \vex(\omega_\times) = \omega.
\end{align}

The wedge, $\wedge$, and vee, $\vee$, operators can be used to transform between matrix and vector representations of the lie algebra, respectively
\begin{align}
&\begin{bmatrix} \xi_R \\ \xi_v \\ \xi_x \end{bmatrix}^\wedge
:= \begin{bmatrix}\begin{matrix} (\xi_R)_\times & \xi_v & \xi_x \end{matrix} \\ \bm{0}_{2\times 5} \end{bmatrix},
\\
&\begin{bmatrix}\begin{matrix} (\xi_R)_\times & \xi_v & \xi_x \end{matrix} \\ \bm{0}_{2\times 5} \end{bmatrix} ^\vee := \begin{bmatrix} \xi_R \\ \xi_v \\ \xi_x \end{bmatrix}.
\end{align}
We also use the vee to denote the matrix representation of a linear group operator. For example, for a group operator $A : \gothg \to \gothg$, the matrix representation, $A^\vee$, satisfies
\begin{align}
(A\circ g)^\vee &= A^\vee g^\vee.
\end{align}

\subsection{Miscellaneous Operators}

The symmetric projector is defined as
\begin{align}
\mathbb{P}_s : \Real{n\times n} \rightarrow \Real{n \times n}, &&& \projsym{A} := \frac{1}{2}(A + A^\top).
\end{align}
We define the matrix operators $\bar{F}$ and $\bar{G} : \Real{3} \rightarrow \Real{5 \times 9}$ as
\begin{align}
\bar{F}(v) &:= \begin{bmatrix} \begin{matrix}-v_\times & \bm{0} & \bm{I}_3 \end{matrix} \\ \bm{0}_{2 \times 9} \end{bmatrix},
\\
\bar{G}(v) &:= \begin{bmatrix} v_\times & \bm{0} & \bm{0} \\ \bm{0} & v^\top & \bm{0} \\ \bm{0} & \bm{0} & v^\top\end{bmatrix}.
\end{align}
For an element $X \in \se_2(3)$, the following identities allow us to transform between the matrix and vector representation of the Lie algebra;
\begin{align}
X \bar{v} &= \bar{F}(v) X^\vee,
\label{eq:identity_F}
\\
X^\top \bar{v} &= \bar{G}(v) X^\vee,
\label{eq:identity_G}
\end{align}
where the bar operator transforms a vector to homogeneous coordinates. Given a vector, $v \in \R^3$,
\begin{align}
\bar{v} &:= \begin{bmatrix}v \\ 0 \\ 1 \end{bmatrix}.
\end{align}
We will also utilise the non-homogeneous representations of $F$ and $G : \Real{3} \rightarrow \Real{3\times 9}$;
\begin{align}
F(v) &= \begin{bmatrix}
-v_\times & \bm{0} & \bm{I}_3
\end{bmatrix},
\\
G(v) &= \begin{bmatrix}
v_\times & \bm{0}
\end{bmatrix}.
\end{align}

\section{Problem Formulation}
\label{sec:problem_formulation}
We consider a single robot operating in free space. We represent the position, $x \in \Real{3}$, orientation, $R \in \SO(3)$, and linear velocity, $v \in \Real{3}$, all with respect to the inertial frame and expressed in the coordinates of the inertial frame.

The kinematics of the vehicle is represented by
\begin{subequations}
\begin{align}
\dot{R} &= R \Omega_\times,
\\
\dot{v} &= R a,
\\
\dot{x} &= v,
\end{align}
\label{eq:kinematics}
\end{subequations}
where $\Omega \in \Real{3}$ is the angular velocity, and $a \in \Real{3}$ is the linear acceleration, both in the coordinates of the body-fixed frame.
We can express the state of the robot, $(R, v, x)$,  as an element of the matrix Lie group $\SE_2(3)$,
\begin{align}
g &:= \begin{bmatrix}
R & v & x \\ \bm{0} & 1 & 0 \\ \bm{0} & 0 & 1
\end{bmatrix}.
\end{align}
The matrix group representation of the kinematics in \eqref{eq:kinematics} is then
\begin{align}
\dot{g} &= g \begin{bmatrix}\begin{matrix} \Omega_\times & a & R^\top v \end{matrix} \\ \bm{0}_{2\times 5} \end{bmatrix}.
\label{eq:kinematics_group}
\end{align}

The robot is equipped with an IMU, which measures linear acceleration, $u_a$, and angular velocity, $u_\omega$, of the body-fixed frame in the coordinates of the body-fixed frame. The measurements are modelled by
\begin{subequations}
\begin{align}
u_\Omega &= \Omega + B_\Omega \delta_\Omega,
\\
u_a &= a + B_a \delta_a,
\end{align}
\label{eq:measurements}
\end{subequations}
where $\delta_\Omega, \delta_a$ are unknown error signals assumed to be zero mean and square integrable with values in $\Real{3}$. $B_\Omega$ and $B_a : \R^3 \to \R^3$ are known linear maps. 

In addition to the IMU, the robot is also equipped with a sensor that measures relative translations to a number of fixed landmarks, $l_i \in \Real{3}, i \in \{1, \ldots, N\}$, in the environment. The measurement signal for each of the landmarks, $y_i(t)$, is modelled by
\begin{subequations}
\begin{align}
y_i(t) &= R^\top(l_i - x) + D \epsilon_i
\\
&= h_i(g) + D \epsilon_i,
\label{eq:landmark_meas}
\end{align}
\end{subequations}
where $\epsilon_i$ is the unknown, zero mean, square integrable measurement error signal with values in $\Real{3}$. $D: \R^3 \to \R^3$ is a known invertible linear map.

Consider the following cost function on the system;
\begin{align}
J_t &= 
\frac{1}{2} m_0(g_0) + \frac{1}{2} \int_{t_0}^t \norm{\delta(\tau)}^2 + \sum_i \norm{\epsilon_i(\tau)}^2 d\tau
\label{eq:cost_functional}
\end{align}
where $\delta =\begin{bmatrix}\delta_\Omega \\ \delta_a \end{bmatrix}$ and $m_0: \SE_2(3) \to \R$ is a bounded smooth function with a unique global minimum.
We define the minimising trajectory $g^*_{[t_0, t]}$ as the trajectory which is compatible with the system kinematics and measurement model and that minimises $J_t$.

The filter estimate, $\hat{g}$, is defined as the terminal point of the minimising trajectory over the time period $[t_0, t]$. More specifically,
\begin{align}
\hat{g}(t) &= g^*_{[t_0, t]}(t).
\end{align}

\section{Abstract Filter}
\label{sec:derivation}
In this section, we present the abstract operator formulation of the filter, drawing on the results from \cite{sacconSecondOrderOptimalMinimumEnergyFilters2016} as the basis for the derivation.

Substituting the measurement model from \eqref{eq:measurements} into \eqref{eq:kinematics_group} gives
\begin{subequations}
\begin{align}
\dot{g} &= g ( \begin{bmatrix} \begin{matrix}(u_{\Omega})_\times & u_a & R^\top v \end{matrix} \\ \bm{0}_{2 \times 5} \end{bmatrix} + \begin{bmatrix} \begin{matrix} (-B_\Omega \delta_\Omega)_\times & -B_a \delta_a & 0 \end{matrix}\\ \bm{0}_{2 \times 5} \end{bmatrix} )
\\
&= g \left( \lambda_t(g, u) + B(\delta(t) )\right)
\label{eq:full_model}
\end{align}
\end{subequations}
where $\lambda_t(g, u)$ is shorthand for $\lambda(g(t), u(t))$.

The model described by \eqref{eq:full_model} and \eqref{eq:landmark_meas}, together with the cost functional \eqref{eq:cost_functional} is a case of the general model and cost functional described by Saccon et al. in \cite{sacconSecondOrderOptimalMinimumEnergyFilters2016}. Because of this, we can directly use the resulting filter equations derived by Saccon et al.

From \cite{sacconSecondOrderOptimalMinimumEnergyFilters2016}, the second-order optimal\footnote{
	The derivation of the filter utilises the value function, $V(g,t)$, which is defined as the minimising cost among all trajectories of \eqref{eq:full_model} in the interval $[t_0, t]$ which reach the state $g \in \SE_2(3)$ at time $t$. Repeatedly differentiating the value function with respect to time provides a set of necessary conditions which define the optimal filter. Ordinarily, this would result in an infinite number of conditions, and so a key step in the filter derivation is to approximate the value function to second order, and assume that all higher-order terms are negligible. This is the reason we refer to the filter as \textit{second-order optimal} \cite{sacconSecondOrderOptimalMinimumEnergyFilters2016}.
} minimum energy estimate, $\hat{g}$, for the state of the system described above is
\begin{align}
\dot{\hat{g}} &= \hat{g} \left( \lambda_t(\hat{g}, u) + K(t) r_t(\hat{g}) \right),
\label{eq:abstract_filter}
\\
\hat{g}(t_0) &= \argmin_g m_0(g).
\end{align}
The residual, $r_t \in \se_2(3)^*$, is given by
\begin{gather}
r_t(\hat{g}) = \sum_i T_eL^*_{\hat{g}} \circ \left[ \left( P_y \circ (y_i - \hat{y}_i) \right) \circ \der h_i(\hat{g}) \right]
\label{eq:r_operator}
\end{gather}
where $P_y =  (D\inv)^* \circ D\inv$ and  $\hat{y}_i = h_i(\hat{g})$.

The gain operator, $K(t) : \se_2(3)^* \rightarrow \se_2(3)$, satisfies the  Riccati equation
\begin{gather}
\begin{multlined}
\dot{K} = A \circ K + K \circ A^* - K \circ E \circ K + B \circ  B^*
\\ - \omega_{Kr} \circ K - K \circ \omega^*_{Kr}
\label{eq:K_dot}
\end{multlined}
\intertext{where}
K(t_0) = X_0\inv, \quad X_0 = T_eL^*_{\hat{g}_0} \circ \Hess m_0(\hat{g}_0) \circ T_eL_{\hat{g}},
\\
A(t) =\der_1 \lambda_t(\hat{g}, u) \circ T_eL_{\hat{g}} - \ad_{\lambda_t(\hat{g}, u)} - T_{\lambda_t(\hat{g}, u)},
\label{eq:A(t)_operator}
\\
\begin{multlined}
E(t) = \sum_i - T_eL^*_{\hat{g}} \circ \Big[ \left( P_y \circ (y_i - \hat{y}_i) \right) ^ {T_{\hat{g}}G} \circ \Hess h_i(\hat{g})
\\- (\der h_i(\hat{g}))^* \circ P_y  \circ \der h_i(\hat{g}) \Big] \circ T_eL_{\hat{g}},
\end{multlined}
\label{eq:E(t)_operator}
\end{gather}
and $Kr$ is shorthand notation for $K r_t(\hat{g})$.

There are a number of possible choices for the connection function  in \eqref{eq:K_dot}, $\omega$, and in this case we select the Cartan (0)-Connection, which is defined as
\begin{align}
\omega^{(0)}(X, Y) &= \frac{1}{2} [X, Y] = \frac{1}{2} \ad_X(Y).
\end{align}

\section{Explicit Filter}
\label{sec:explicit_filter}
The filter equations presented in Section \ref{sec:derivation} represent abstract operations on a general Lie group. In this section we present one of the main contributions of this work, which is to determine the explicit matrix representations of these operators, specialised to the $\SE_2(3)$ Lie group. Choosing the standard basis for the Lie algebra, we detail the matrix representations of the relevant operators in the proceeding 4 lemmas. These culminate in an explicit representation for the entire filter, as detailed in Theorem \ref{thm:explicit_filter}.
\begin{lemma}
	\label{lem:adjoint}
	Consider the adjoint representation of $\se_2(3)$, $\ad_\xi$. This operator on the Lie algebra can be equivalently represented by the matrix operator
	\begin{align}
	\ad_\xi^\vee &= \begin{bmatrix}
	\xi_{R\times} & \bm{0} & \bm{0} \\
	\xi_{v\times} & \xi_{R\times} & \bm{0} \\
	\xi_{x\times} & \bm{0} & \xi_{R\times}
	\end{bmatrix}.
	\end{align}
	\begin{proof}
		The adjoint operator for a matrix Lie group is equivalent to the matrix commutator;
		\begin{align}
		\ad_\xi \gamma = [\xi, \gamma] = \xi \gamma - \gamma \xi.
		\end{align}
		We then expand the matrix multiplication using the group representation in \eqref{eq:def_se_2(3)} and use the identities on the cross product,
		\begin{align}
		\omega_\times \phi &= - \phi_\times \omega,
		\\
		[\omega_\times, \phi_\times] &= (\omega_\times \phi)_\times,
		\end{align}
		for $\omega, \phi \in \Real{3}$.
	\end{proof}
\end{lemma}	

\begin{lemma}
	\label{lem:A(t)}
Consider the operator $A(t) : \se_2(3) \to \se_2(3)$ defined in  \eqref{eq:A(t)_operator}.  Let $A^\vee(t) \in \R^{9 \times 9}$ denote the matrix representation of $A(t)$.
Then
	\begin{align}
	A^\vee(t) &:= \begin{bmatrix}
	-u_{\Omega\times} & 0 & 0\\
	-u_{a\times} &  -u_{\Omega\times} & 0 \\
	0 & I &  -u_{\Omega\times}
	\end{bmatrix}.
	\end{align}

	\begin{proof}
		We determine the matrix representation of the first term of \eqref{eq:A(t)_operator} by taking the directional derivative in an arbitrary direction, $gX \in T_{\hat{g}} \SE_2(3)$,
		\begin{align}
		\der_1 \lambda_t(\hat{g}, u) \circ \hat{g}X &= \left(\begin{bmatrix}
		0 & 0 & 0 \\ 0 & 0 & 0 \\ (R^\top v)_\times & I & 0
		\end{bmatrix}
		X^\vee \right)^\wedge.
		\end{align}
		Considering the direction was arbitrary, this gives
		\begin{align}
		\left(\der_1 \lambda_t(\hat{g}, u) \circ T_eL_{\hat{g}}\right)^\vee &= \begin{bmatrix}
		0 & 0 & 0 \\ 0 & 0 & 0 \\ (R^\top v)_\times & I & 0
		\end{bmatrix}
		\end{align}
		as the matrix representation. The second term of \eqref{eq:A(t)_operator} is evaluated by using Lemma \ref{lem:adjoint} and the final term is zero as the (0)-connection has trivial torsion, i.e. $T(X, Y) = 0$. The result then follows.
	\end{proof}
\end{lemma}

\begin{lemma}
	\label{lem:E(t)}
	Consider the operator $E(t) : \se_2(3) \to \se_2(3)^*$ defined in  \eqref{eq:E(t)_operator}.  Let $E^\vee(t) \in \R^{9 \times 9}$ denote the matrix representation of $E(t)$.
	Then
	\begin{multline}
	E^\vee(t) = \sum_i \Big[ - \projsym{F(\hat{y}_i)^\top G\big( P_y (y_i - \hat{y}_i) \big)} \\+ F(\hat{y}_i)^\top P_y  F(\hat{y}_i) \Big].
	\end{multline}
	
	\begin{proof}
		We apply the operator in \eqref{eq:E(t)_operator} to two arbitrary elements of $\se_2(3)$, $X$ and $Y$, and use the identity in \eqref{eq:exponential_functor}, which gives
		\begin{multline}
		[E(t)\circ X]\circ Y =
		\\
		 \sum_i \Big[ -  \left( P_y \circ (y_i - \hat{y}_i) \right) \circ \left[\Hess h_i(\hat{g}) \circ \hat{g} X \right] \circ \hat{g} Y
		\\
		 + \left[ \left[ (\der h_i(\hat{g}))^* \circ P_y  \circ \der h_i(\hat{g}) \right] \circ \hat{g} X \right] \circ \hat{g} Y \Big].
		 \label{eq:E(t)_X_Y}
		\end{multline}
		
		The Hessian operator, in homogeneous coordinates, is given by
		\begin{align}
		\begin{split}
		[\Hess &~ \bar{h}_i(\hat{g})  \circ \hat{g}X] \circ \hat{g}Y \\ &= \der (\der \bar{h}_i(\hat{g}) \circ \hat{g}Y) \circ \hat{g}X  - \der \bar{h}_i(\hat{g}) \circ \hat{g} \omega_X(Y)
		\end{split}
		\\
		&= \frac{1}{2} \left(XY \bar{h}_i(\hat{g}) + YX \bar{h}_i(\hat{g})\right).
		\end{align}
		Substituting the Hessian operator into \eqref{eq:E(t)_X_Y} and evaluating the derivatives in the second term gives
		\begin{multline}
		[E(t)\circ X]\circ Y =  \sum_i \Big[ - \frac{1}{2} \left\langle  P_{\bar{y}}  (\bar{y}_i - \bar{\hat{y}}_i) , XY \bar{\hat{y}}_i + YX \bar{\hat{y}}_i \right\rangle
		\\
		 + \left\langle P_{\bar{y}} , X \bar{\hat{y}}_i   Y \bar{\hat{y}}_i \right \rangle \Big]
		\end{multline}
		where $\hat{y}_i = h_i(\hat{g})$.
		
		We then utilise the identities in \eqref{eq:identity_F}, \eqref{eq:identity_G} to isolate the $X$ and $Y$ terms, giving
		\begin{multline}
		[E(t)\circ X]\circ Y =\\ \sum_i \Big[ - \frac{1}{2}  (Y^{\vee}) ^\top \bar{F}(\hat{y}_i)^\top \bar{G}\left( P_y  (y_i - \hat{y}_i) \right)  X^\vee
		\\
		 \shoveright{ -\frac{1}{2} (Y^\vee)^\top \bar{G}\left( P_y (y_i - \hat{y}_i) \right)^\top  \bar{F}(\hat{y}_i) X^{\vee} }
		 \\
		 + (Y^\vee)^\top \bar{F}(\hat{y}_i)^\top P_{\bar{y}}  \bar{F}(\hat{y}_i)  X^\vee \Big] .
		\end{multline}
		From this, we convert back from homogeneous coordinates and the lemma follows.
	\end{proof}
\end{lemma}

\begin{lemma}
	\label{lem:residual}
	Consider the operator $r_t : \se_2(3) \to \R$ defined in  \eqref{eq:r_operator}.  Let $r_t^\vee \in \R^{9}$ denote the matrix representation of $r_t$.
	Then
	\begin{align}
	r_t^\vee(\hat{g}) &= \sum_i  (y_i - \hat{y}_i)^\top P_y F(\hat{y}_i).
	\end{align}
	\begin{proof}
		Similarly to Lemma \ref{lem:E(t)}, we apply the operator \eqref{eq:r_operator} to an arbitrary element $X \in \se_2(3)$ which gives
		\begin{align}
		r_t(\hat{g}) \circ X &= \sum_i \left\langle P_y \circ (y_i - \hat{y}_i) ,  \bm{d}h_i(\hat{g}) \circ \hat{g}X \right \rangle.
		\end{align}
		We then evaluate the derivative and apply the identity from \eqref{eq:identity_F} and the lemma follows.
	\end{proof}
\end{lemma}

\begin{theorem}
	\label{thm:explicit_filter}
	The explicit matrix representation of the second-order optimal minimum energy filter is given by
	 \begin{align}
	 \dot{\hat{g}} &= \hat{g} \left( \lambda_t(\hat{g}, u) + K(t) r_t(\hat{g}) \right),
	 \\
	 \dot{K} &= \projsym{2 A^\vee K - \ad_{Kr}^\vee K} - KE^\vee K + B^\vee (B^\vee )^\top
	 \end{align}
	 \begin{proof}
	 	This follows as a consequence of Lemmas \ref{lem:A(t)}, \ref{lem:E(t)}, and \ref{lem:residual}.
	 \end{proof}
\end{theorem}

\section{Implementation}
\label{sec:discretisation}
There are a number of considerations that need to be made when implementing the filter on a physical system. Primarly, we must propose a method of discretising the continuous-time differential equations and also consider the effect of differing sensor update rates. In this section, we propose one possible way of discretising the filter given the real-world constraints of the system.

Modern IMU sensors have sufficiently high fixed sample rates that we can simply numerically integrate the terms in the differential equation relating to the IMU measurements. However, we consider that the landmark sensor information may be available at a much slower rate, and potentially intermittently. Thus, for the terms in the differential equation relating to the landmark sensor, we perform a discrete update step. This approach is similar to that of Zamani and Trumpf in \cite{zamaniDiscreteUpdatePose2019}, but we aim to find a discretisation of the continuous time equations, rather than directly derive the discrete update equations.

In discretising the filter, we will find that it is simpler to work with the inverse form of the Riccati equation for the gain matrix, $P = K \inv$, which gives
\begin{gather}
\dot{P} = - \projsym{2PA^\vee - P \ad^\vee_{P\inv r} } + E^\vee - PB^\vee (B^\vee)^\top P.
\label{eq:dot_P}
\end{gather}

Considering just the IMU measurements, we apply the Lie Group Euler method to propagate the state estimate forwards in time from $t$ to $t + \Delta t$, where $\Delta t$ is the time between successive IMU measurements. We also apply the standard Euler method to the gain matrix $P$, excluding the terms that are dependent on the landmark measurement. This gives
\begin{align}
\hat{g}(t + \Delta t) =& \hat{g}(t) \cdot \exp(\Delta t \cdot \lambda_t(\hat{g}, u)),
\\
\begin{split}
P(t + \Delta t) =&P(t)  -2 \Delta t  \cdot \projsym{P(t)A^\vee(t)} \\
&- \Delta t \cdot P(t)B^\vee (B^\vee)^\top P(t),
\
\end{split}
\label{eq:P_predict}
\end{align}
where $\exp$ is the matrix exponential.

When landmark measurements are available at some time, $t$, we perform a discrete update step, updating the state and gain matrix from $\hat{g}(t)$ and $P(t)$ to $\hat{g}(t^+)$ and $P(t^+)$ respectively,
\begin{gather}
\hat{g}(t^+) = \hat{g}(t) \cdot \exp\left( \alpha \left[ P(t^+)\inv r^\vee_t(\hat{g}) \right]^\wedge \right),
\\
P(t^+) = P(t) + \alpha E^\vee(t) + \alpha \projsym{P(t) \ad^\vee_{P(t)\inv r_t^\vee(\hat{g})}}.
\label{eq:P_update}
\end{gather}
It is necessary to apply a gain, $\alpha$, to account for the disparity in frequency between landmark and velocity measurements. Consider that, if landmark measurements were available at the same frequency as velocity measurements, then directly taking the Euler integration of \eqref{eq:dot_P} would result the same equations as \eqref{eq:P_predict} and \eqref{eq:P_update} with $\alpha = \Delta t$. Additionally, if at a given time step only a subset of landmark measurements are available, then the summands in $E(t)$ and $r_t$ corresponding to other landmarks become zero. For example, this might occur in a real-world scenario if a landmark is out of range or obscured by an obstacle.

\section{Simulation}
\label{sec:simulations}

In this section, we demonstrate an implementation of the discrete-time minimum energy filter from section \ref{sec:discretisation}. We consider a single UAV operating in free space on a trajectory with a constant angular and linear velocity. The filter is initialised with a displaced state estimate and an initial value for $P = \diag (10^{-3}, 10^{-3}, 10^{-3}, 3,3,3,5,5,5)$.

We model the IMU with an update rate of $1000$ Hz and model the noise with a normal distribution, $\delta \sim \mathcal{N}(0, 1)$. We select the gains $B_\Omega = 0.1 \bm{I}_3$ and  $B_a = 0.1 \bm{I}_3$.

We position 4 fixed landmarks in the environment at various points. The landmark sensor measures the relative position of all landmarks at a fixed rate of $10$ Hz, with the sensor error drawn from a random distribution, $\epsilon \sim \mathcal{N}(0, 1)$ and the gain terms selected as $D = 0.5 \bm{I}_3$, $\alpha = 0.1$.

\begin{figure}
	\centering
	\includegraphics[width=1\linewidth]{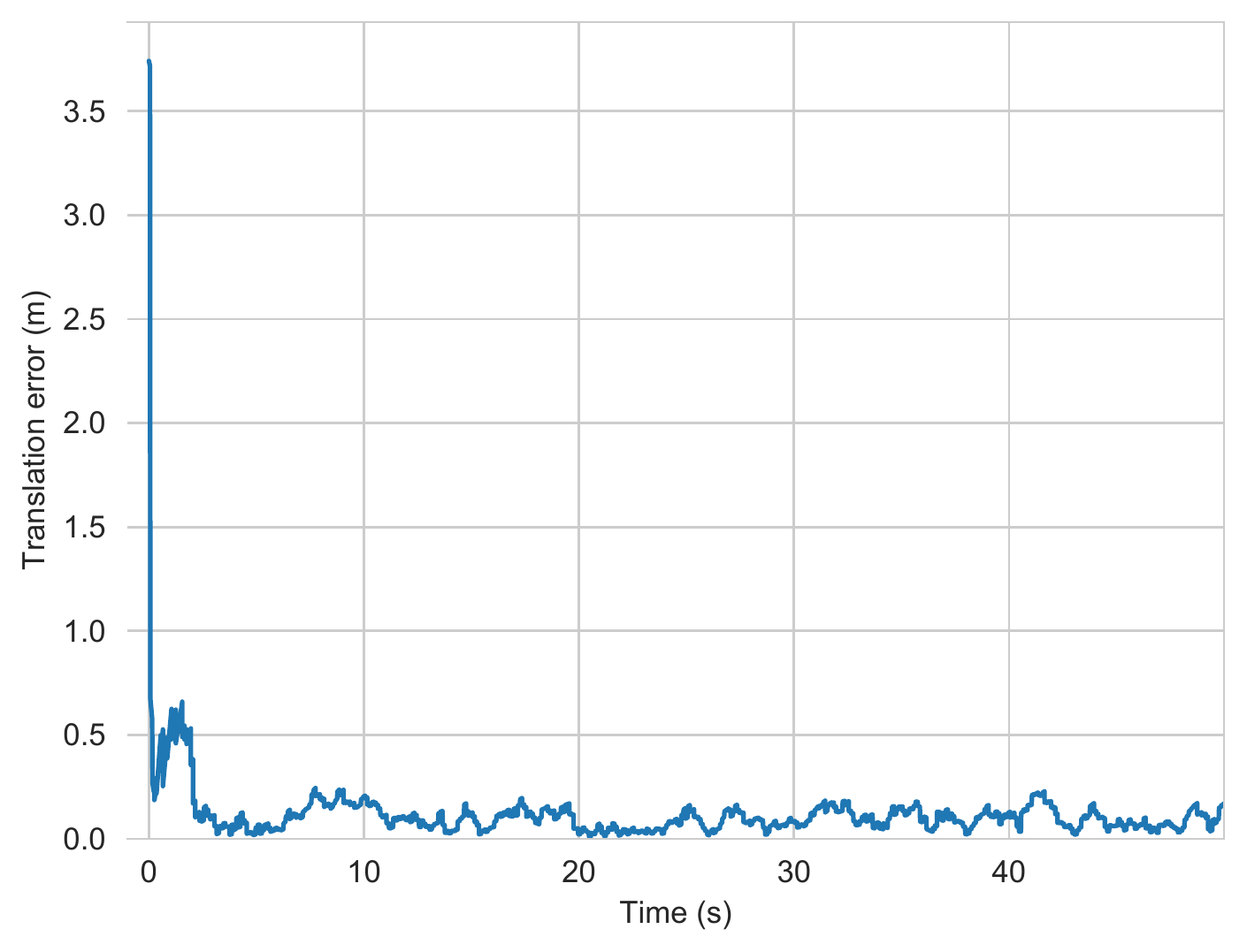}
	\caption{Translation error ($l^2$-norm) of filter estimate for a typical simulation}
	\label{fig:translation_error}
\end{figure}

\begin{figure}
	\centering
	\includegraphics[width=1\linewidth]{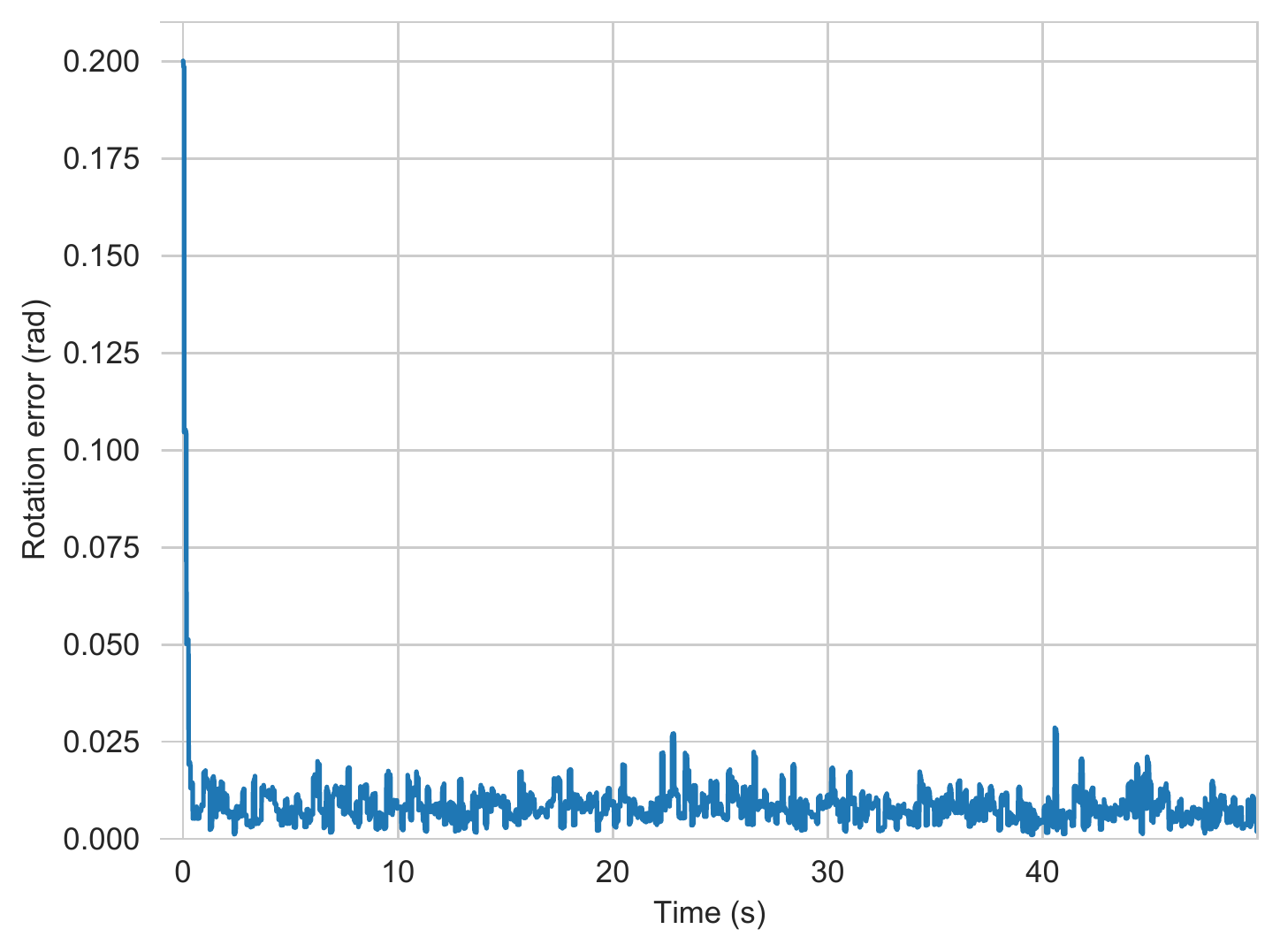}
	\caption{Rotation error of filter estimate for a typical simulation}
	\label{fig:rotation_error}
\end{figure}

The results of the simulation are shown in Figures \ref{fig:translation_error} and \ref{fig:rotation_error}. The two graphs show that the filter is able to relocalise after a large initialisation error and maintain a consistently low error in the state estimate. The filter converges from an initial translation error of 3.8m and an initial rotation error of 0.2 radians to an average error of 0.12m and 0.009 radians respectively.

\section{Conclusion}
\label{sec:conclusion}

This paper proposes a minimum energy filter for pose estimation of an aerial vehicle based on IMU and landmark measurements. 
The key contribution of the paper lies in specialising the results of 
\cite{sacconSecondOrderOptimalMinimumEnergyFilters2016} to the specific case of the extended special Euclidean group $\SE_2(3)$  \cite{barrauEKFSLAMAlgorithmConsistency2016} and providing an explicit matrix representation of the filter.  
We also propose an asynchronous discrete-time implementation to fuse high bandwidth IMU with low bandwidth discrete-time landmark measurements typical of real-world scenarios.

\balance
\bibliographystyle{IEEEtran}
\bibliography{references_bibtex}             

\end{document}

%% file: macros.tex
\newcommand{\g}{\mathfrak{g}}
\newcommand{\SE}{\mathrm{SE}}
\newcommand{\se}{\mathfrak{se}}
\newcommand{\SO}{\mathrm{SO}}
\newcommand{\so}{\mathfrak{so}}

\newcommand{\inv}{^{\,\text{-}1}}

\newcommand*{\Real}[1]{\ensuremath{\mathbb{R}^{#1}}}

\newcommand\norm[1]{\ensuremath{\left\Vert #1 \right\Vert}}

\newcommand\projsym[1]{\ensuremath{\mathbb{P}_s\left( #1 \right)}}

\DeclareMathOperator*{\argmin}{arg\,min}
\DeclareMathOperator{\diag}{diag}

\DeclareMathOperator{\vex}{vex}
\DeclareMathOperator{\ad}{ad}
\DeclareMathOperator\Hess{Hess}
\bmdefine{\der}{\mathrm{d}}

%% file: notation_defs.tex
\providecommand{\R}{\mathbb{R}}

\providecommand{\SO}{\mathbf{SO}}

\providecommand{\SE}{\mathbf{SE}}

\providecommand{\so}{\mathfrak{so}}
\providecommand{\se}{\mathfrak{se}}

\renewcommand{\sl}{\mathfrak{sl}}

\providecommand{\gothg}{\mathfrak{g}}

\providecommand{\Hess}{\mathrm{Hess}}

\usepackage{accents}
\makeatletter
\providecommand{\scirc}{%
    \hbox{\fontfamily{lmr}\fontsize{.4\dimexpr(\f@size pt)}{0}\selectfont{\raisebox{-0.34ex}[0ex][-0.34ex]{$\circ$}}}}

\makeatother

\mathchardef\mhyphen="2D

\providecommand{\etal}{\textit{et al.}~}

